\documentclass[prd,aps,floats,preprintnumbers,preprint]{revtex4}

\usepackage{graphicx}
 
\newcommand{\be}{\begin{equation}}
\newcommand{\ee}{\end{equation}}
\newcommand{\bea}{\begin{eqnarray}}
\newcommand{\eea}{\end{eqnarray}}

\begin{document}

\preprint{YITP-09-109}
\preprint{KUNS-KUNS 2248}

\setlength{\unitlength}{1mm}

\title{Mimicking the cosmological constant for more than one observable with large scale inhomogeneities}

\author{Antonio Enea Romano}
\affiliation{
$^1$Leung Center for Cosmology and Particle Astrophysics, National Taiwan University, Taipei 10617, Taiwan, R.O.C.\\
$^2$Yukawa Institute for Theoretical Physics, Kyoto University,
Kyoto 606-8502, Japan
\\
}


\begin{abstract}
Assuming the definition of the inversion problem (IP) as the exact matching of the terms in the low redshift expansion of cosmological observables calculated for different cosmological models, we solve the IP for $D_L(z)$ and the redshift spherical shell mass density $mn(z)$ for a central observer in a LTB space without cosmological constant and a generic $\Lambda CDM $ model. We show that the solution of the IP is unique, corresponds to a matter density profile which is not smooth at the center and that the same conclusions can be reached expanding self-consistently to any order all the relevant quantities.

On contrary to the case of a single observable inversion problem, it
is impossible to solve the IP (LTB vs. $\Lambda$CDM)
 for both $mn(z)$ and $D_L(z)$ while setting one the two functions $k(r)$ or $t^b(r)$ to zero, even allowing not smooth matter profiles. Our conclusions are general, since they are exclusively based on comparing directly physical observables in redshift space, and don't depend on any special ansatz or restriction for the functions defining a LTB model.

\end{abstract}

\maketitle

\section{Introduction}
High redshift luminosity distance measurements \cite{Perlmutter:1999np,Riess:1998cb,Tonry:2003zg,Knop:2003iy,Barris:2003dq,Riess:2004nr} and
the WMAP measurements \cite{WMAP2003,Spergel:2006hy} of cosmic
microwave background (CMB) interpreted in the context of standard 
FLRW cosmological models have strongly disfavored a matter dominated universe,
 and strongly supported a dominant dark energy component, giving rise
to a positive cosmological acceleration.
As an alternative to dark energy, it has been 
proposed \cite{Nambu:2005zn,Kai:2006ws}
 that we may be at the center of an inhomogeneous isotropic universe without cosmological constant described by a Lemaitre-Tolman-Bondi (LTB)  solution of Einstein's field 
equations, where spatial averaging over one expanding and one contracting 
region is producing a positive averaged acceleration $a_D$, but it has been shown how spatial averaging can give rise to averaged quantities which are not observable \cite{Romano:2006yc}.

Another more general approach to map luminosity distance as a function of
 redshift $D_L(z)$ to LTB models has been recently
 proposed \cite{Chung:2006xh,Yoo:2008su},
 showing that an inversion method can be applied successfully to 
reproduce the observed $D_L(z)$.   

Interesting analysis of observational data in inhomogeneous models without dark energy and of other theoretically related problems is given for example in \cite{Alexander:2007xx,Alnes:2005rw,GarciaBellido:2008nz,GarciaBellido:2008gd,GarciaBellido:2008yq,February:2009pv,Uzan:2008qp,Quartin:2009xr,Quercellini:2009ni}.
Our definition of the inversion problem is purely mathematical, and consists in matching exactly the terms of low redshift expansion for the relevant cosmological observables for different models. Following this procedure we determine the LTB models which are mimicking at low redshift the $\Lambda CDM$ theoretical predictions.

It is important to observe that according to our definition of the IP we look for the LTB models which match the coefficients of the redshift expansion of the observables corresponding to the best fit flat $\Lambda CDM$ models, but this kind of approach is not completely rigorous since while these models depend only on the two independent parameters $H_0,\Omega_M$, LTB models have a higher number of parameters, implying that they could actually provide an even better fit of experimental data. 
From an observational point of view it would in fact be more important to confront directly with actual experimental data rather than reproduce the best fit theoretical $\Lambda CDM$ model, so our conclusions should be considered keeping this limitation in mind, and we leave to a future work such a direct analysis of experimental data.
Our definition of the IP appears nevertheless the most natural one from a mathematical point of view since it consists in matching the theoretical predictions of different models within the range of validity of the Taylor expansion, and the formulae derived could actually be used even for the purpose of low-redshift data fitting, since they have the advantage of not being dependent on any functional ansatz.
We could for example obtain contour plots for the coefficients of the expansion of the functions $k(r)$ , in the same way we obtain for $\Omega_{\Lambda}$ for example.

In \cite{Clifton:2008hv} it is also given some direct evidence that the smoothness of the inhomogeneity profile is important to allow to distinguish these models from $\Lambda CDM$, which is consistent with our conclusion that the solution of the inversion problem is possible only if we allow a not smooth radial matter profile, while smooth models \cite{Romano:2009ej} can be distinguished from $\Lambda CDM$, at least in principle, without any ambiguity.

The main point is that the luminosity distance is in general sensitive 
to the geometry of the space through which photons are propagating along
null geodesics, and therefore arranging appropriately the geometry of 
a given cosmological model it is possible to reproduce a given $D_L(z)$.
For FLRW models this corresponds to the determination of $\Omega_{\Lambda}$
 and $\Omega_m$ and for LTB models it allows to determine the 
functions $E(r),M(r),t_b(r)$.

Another observable which could be used to constraint LTB models is the redshift spherical shell mass $mn(z)$ \cite{Mustapha:1998jb}, which has been recently calculated  \cite{Romano:2009qx} for a central observer up to the fifth order in the redshift, and can be generalized \cite{Romano:2007zz} to the more observationally related $E_{RSS}(z)$, the redshift spherical shell energy.

Since $mn(z)$ is not yet available observationally, we don't know if it is in agreement with $\Lambda CDM$ models. It could be possible that only LTB models can fit the observed $mn(z)$, in which case it would allow to distinguish between LTB and $\Lambda CDM$ models. 
If on the contrary $mn(z)$ is in agreement with $\Lambda CDM$ predictions, we could still not distinguish between them, as shown numerically in \cite{{Celerier:2009sv}}, and analytically in this paper. In this case there would be only one LTB model able to exactly mimick the mathematical $\Lambda CDM$ prediction for both $mn(z)$ and $D_L(z)$ and it would corresponds to a matter density profile not smooth at the center.

In this paper we derive for the first time in the FRLW gauge the general formula to third order in redshift for the luminosity distance for a central observer in a matter dominated LTB space without imposing the conditions for a smooth matter distribution.

We then calculate the low redshift expansion of $mn(z)$ and $D_L(z)$ for a flat $\Lambda CDM$ model and show how it possible for both these observables to mimick with a matter dominated LTB model the effects of the cosmological constant in a flat homogeneous space, that the solution is unique and corresponds to a matter profile which is not smooth at the center.

Our results are general since we don't use any special ansatz for the functions defining LTB models, and the arguments about the number of free parameters defining the LTB models constrained by observations are independent of the value of the cosmological constant.

We need three functions to define a LTB solution, but because of the invariance under general coordinate transformations, only two of them are really independent. This implies that two observables are in principle sufficient to solve the IP of mapping observations to a specific LTB model, for example the luminosity distance $D_L(z)$ and the redshift spherical shell mass $m(z)n(z)=mn(z)$.

As observed by \cite{Celerier:2009sv}, there has been sometime some confusion about the general type of LTB models which could be used to explain cosmological observations, so it is important to stress that without restricting the attention on models with homogeneous big bang, a void is not necessary to explain both $D_L(z)$ and $m(z)n(z)=mn(z)$ without the cosmological constant, but we will prove that this is only possible for LTB models that are not smooth at the center.
 

\section{Lemaitre-Tolman-Bondi (LTB) Solution\label{ltb}}
Lemaitre-Tolman-Bondi  solution can be
 written as \cite{Lemaitre:1933qe,Tolman:1934za,Bondi:1947av}
\begin{eqnarray}
\label{eq1} %
ds^2 = -dt^2  + \frac{\left(R,_{r}\right)^2 dr^2}{1 + 2\,E}+R^2
d\Omega^2 \, ,
\end{eqnarray}
where $R$ is a function of the time coordinate $t$ and the radial
coordinate $r$, $R=R(t,r)$, $E$ is an arbitrary function of $r$, $E=E(r)$
and $R,_{r}=\partial R/\partial r$.

Einstein's equations give
\begin{eqnarray}
\label{eq2} \left({\frac{\dot{R}}{R}}\right)^2&=&\frac{2
E(r)}{R^2}+\frac{2M(r)}{R^3} \, , \\
\label{eq3} \rho(t,r)&=&\frac{2 M,_{r}}{R^2 R,_{r}} \, ,
\end{eqnarray}
with $M=M(r)$ being an arbitrary function of $r$ and the dot denoting
the partial derivative with respect to $t$, $\dot{R}=\partial R(t,r)/\partial t$.



We can introduce the variables
\begin{equation}
 A(t,r)=\frac{R(t,r)}{r},\quad k(r)=-\frac{2E(r)}{r^2},\quad
  \rho_0(r)=\frac{6M(r)}{r^3} \, ,
\end{equation}
so that  Eq.\ (\ref{eq1}) and the Einstein equations
(\ref{eq2}) and (\ref{eq3}) are written in a form 
similar to those for FLRW models,
\begin{equation}
\label{eq6} ds^2 =
-dt^2+A^2\left[\left(1+\frac{A,_{r}r}{A}\right)^2
    \frac{dr^2}{1-k(r)r^2}+r^2d\Omega_2^2\right] \, ,
\end{equation}
\begin{eqnarray}
\label{eq7} %
\left(\frac{\dot{A}}{A}\right)^2 &=&
-\frac{k(r)}{A^2}+\frac{\rho_0(r)}{3A^3} \, ,\\
\label{eq:LTB rho 2} %
\rho(t,r) &=& \frac{(\rho_0 r^3)_{, r}}{3 A^2 r^2 (Ar)_{, r}} \, .
\end{eqnarray}
The solution can now be written as
\begin{eqnarray}
\label{LTB soln2 R} a(\eta,r) &=& \frac{\rho_0(r)}{6k(r)}
     \left[ 1 - \cos \left( \sqrt{k(r)} \, \eta \right) \right] \, ,\\
\label{LTB soln2 t} t(\eta,r) &=& \frac{\rho_0(r)}{6k(r)}
     \left[ \eta -\frac{1}{\sqrt{k(r)}} \sin
     \left(\sqrt{k(r)} \, \eta \right) \right] + t_{b}(r) \, ,
\end{eqnarray}
where $\eta \equiv \tau\, r = \int^t dt'/A(t',r) \,$ and $A(t(\eta,r),r)=a(\eta,r)$ and $t_{b}(r)$ is another arbitrary function of $r$, called the bang function,
which corresponds to the fact that big-bang/crunches can happen at different
times. This inhomogeneity of the location of the singularities is one of
the origins of the possible causal separation \cite{Romano:2006yc} between 
the central observer and the spatially averaged region for models
 with positive $a_D$.

In the rest of paper we will use this last set of equations.
Furthermore, without loss of generality, we may set 
the function $\rho_0(r)$ to be a constant,
 $\rho_0(r)=\rho_0=\mbox{constant}$, corresponding to the choice of coordinates in which $M(r)\propto r^3$, and we will call this, following \cite{Tanimoto:2007dq}, the FLRW gauge.


\section {Geodesic equations}

We will adopt the same method developed in \cite{Romano:2009xw} to find the null geodesic equations in the coordinates $(\eta,t)$, but here instead of integrating numerically the differential equations we will find a local expansion of the solution around $z=0$ corresponding to the point $(t_0,0)\equiv(\eta_0,0)$, where $t_0=t(\eta_0,0)$.
We will indeed slightly change notation to emphasize the fully analytical r.h.s. of the equations obtained in terms of $(\eta,t)$, on the contrary of previous versions of the light geodesic equations which require some numerical calculation of $R(t,r)$ from the Einstein's equation(\ref{eq2}). 

For this reason this formulation is particularly suitable for the derivation of analytical results.

The luminosity distance for a central observer in a LTB space 
as a function of the redshift is expressed as
\be
D_L(z)=(1+z)^2 R\left(t(z),r(z)\right)
=(1+z)^2 r(z)a\left(\eta(z),r(z)\right) \,,
\ee
where $\Bigl(t(z),r(z)\Bigr)$ or $\Bigl((\eta(z),r(z)\Bigr)$
is the solution of the radial geodesic equation
as a function of the redshift.

The past-directed radial null geodesic is given by
\bea
\label{geo1}
\frac{dT(r)}{dr}=f(T(r),r) \,;
\quad
f(t,r)=\frac{-R_{,r}(t,r)}{\sqrt{1+2E(r)}} \,.
\eea
where $T(r)$ is the time coordinate along the null radial geodesic as a function of the the coordinate $r$.

From the implicit solution, we can write 
\bea
T(r)=t(U(r),r) \\
\frac{dT(r)}{dr}=\frac{\partial t}{\partial \eta} \frac{dU(r)}{dr}+\frac{\partial t}{\partial r} 
\eea
where $U(r)$ is the $\eta$ coordinate along the null radial geodesic as a function of the the coordinate $r$.
The geodesic equations for the coordinates $(\eta,r)$ can be written as,
\bea
\label{geo3}
\frac{d \eta}{dz}
&=&\frac{\partial_r t(\eta,r)-F(\eta,r)}{(1+z)\partial_{\eta}F(\eta,r)}=p(\eta,r) \,,\\
\label{geo4}
\frac{dr}{dz}
&=&-\frac{a(\eta,r)}{(1+z)\partial_{\eta}F(\eta,r)}=q(\eta,r) \,, \\
F(\eta,r)&=&-\frac{1}{\sqrt{1-k(r)r^2}}\left[\partial_r (a(\eta,r) r)
+\partial_{\eta} (a(\eta,r) r) \partial_r \eta\right]  \, , 
\eea
where $\eta=U(r(z))$ and $F(\eta,r)=f(t(\eta,r),r)$.
It is important to observe that the functions $p,q,F$ have an explicit analytical form which can be obtained from $a(\eta,r)$ and $t(\eta,r)$ as shown below.
 
The derivation of the implicit  solution $a(\eta,r)$ is based on the use of 
the conformal time variable $\eta$, which by construction 
satisfies the relation,
\be
\frac{\partial\eta(t,r)}{\partial t}=a^{-1} \,.
\ee

This means
\bea
t(\eta,r)
&=&t_b(r)+\int^{\eta}_{0}a(\eta^{'},r) d\eta^{'} \, ,
\\
dt&=&a(\eta,r)d\eta+\left(\int^{\eta}_{0}
\frac{\partial a(\eta^{'},r)}{\partial r} d\eta^{'}+t_b^{'}(r)\right) dr \,,
\eea
In order to use the analytical solution we need to find an analytical expression for $F$ and $F_{,\eta}$.

 This can always be done by using
\bea
&& \frac{\partial}{\partial t}=a^{-1}{\frac{\partial}{\partial \eta}} \\
&&\partial_r t(\eta,r)=
\frac{ \rho_0 \, k'(r)}{12 k(r)^{5/2}} 
\left[
3 \sin{ \left( \eta\sqrt{k(r)} \right) }
-\eta \left( 2+\cos{ \left(\eta\sqrt{k(r)}\right) } \sqrt{k(r)} \right)
\right]+t_b'(r)  \, , \\
&&\partial_r \eta=
-a(\eta,r)^{-1}\partial_r t  \, 
\eea 
In this way the coefficients of equations (\ref{geo3}) and (\ref{geo4}) are 
fully analytical, which is a significant improvement over previous 
approaches.

\section{Calculating $D_L(z)$ and $mn(z)$}

Here we will not give the formulae in terms of $\eta_0$ and trigonometric functions, since they are rather complicated and not relevant to the scope of this paper, but rather introduce the following quantities:

\bea
a_0=a(\eta_0,0)=\frac{\tan(\frac{\sqrt{k_0}\eta_0}{2})^2 \rho_0}{3 k_0 (\tan(\frac{\sqrt{k_0}\eta_0}{2})^2+1)}, \label{a0}\\
H_0=\frac{3 k_0^{3/2} \left(\tan(\frac{\sqrt{k_0}\eta_0}{2})^2+1\right)}{\tan(\frac{\sqrt{k_0}\eta_0}{2})^3 \rho_0}, \label{H0}\\
q_0=\frac{1}{2} \left(\tan(\frac{\sqrt{k_0}\eta_0}{2})^2+1\right)\label{q0}, 
\eea

where we have used
\bea
H_0 & = &\frac{\dot{a}(t_0,0)}{a(t_0,0)}, \\
q_0 &= -&\frac{\ddot{a}(t_0,0)a(t_0,0)}{\dot{a}(t_0,0)^2}.
\eea


The derivative respect to $t$ is denoted with a dot, and is calculated using the analytical solution $a(\eta,r)$ and the derivative respect to $\eta$ :
\be
\dot{a}=\partial_{t}a=\partial_{\eta}a\, a^{-1}.
\ee

In deriving the above equations we have expressed all the relevant quantities in terms of $\tan(\frac{\sqrt{k_0}\eta_0}{2})$
using standard trigonometric identities such as $\cos(x)=\frac{1-\tan^2(x/2)}{1+\tan^2(x/2)}$. This has the advantage to make formulae valid for any sign of $k_0$ by analytical continuation and to obtain more compact expressions, which are easier to simplify. Without such an approach the results of the calculations can in fact become quite cumbersome and difficult to treat even with a mathematical software such as MATHEMATICA, which was actually used to derive the results presented in this paper.
For this reason a series of simplifying routines have been developed based on the simplifying procedure of recursively re-expressing any trigonometric function in terms of $\tan(\frac{\sqrt{k_0}\eta_0}{2})$, allowing to reduce to a purely algebraic operation the final simplification.

As an extra check all the results have been independently derived using an alternative method based on the local expansion of the solution of the Einstein's equations in terms of the variables $(t,r)$, but we will report this in a separate paper in preparation about the deceleration parameter in LTB models.

Another convention we will follow in the rest of the paper will be to express everywhere $k_0,\rho_0,\eta_0$ in terms of $H_0,a_0,q_0$ by inverting equations (\ref{a0}-\ref{q0}):
 
\bea
k_0 & = &a_0^2 H_0^2 \tan^2(\frac{\sqrt{k_0}\eta_0}{2}), \\
\rho_0 &= & 3 a_0^3 H_0^2 [1+\tan^2(\frac{\sqrt{k_0}\eta_0}{2})] ,\\
\eta_0 & = & \frac{2 \arctan{\sqrt{2 q_0-1}}}{a_0 H_0 \tan(\frac{\sqrt{k_0}\eta_0}{2})}.
\eea

We may actually set $a_0=1$ by choosing an appropriate system of units, but we will leave it in order to clearly show the number of independent degrees of freedom of the problem and to emphasize the difference with $\Lambda CDM$ models.

Expanding the r.h.s. of the geodesics equation we can easily integrate the corresponding polynomial $q(z),p(z)$, to get $r(z)$ and $\eta(z)$.
It can be shown that in order to obtain $D_L(z)$ to the third order and $mn(z)$ the fourth we need to expand $r(z)$ to the third order and $\eta(z)$ to the second.
In this paper we will try to solve the IP  using the expansion:

\bea
k(r)&=&k_0+k_1 r+k_2 r^2+ ..\\
t_b(r)&=&t^b_0+t^b_1 r+t^b_2 r^2+.. \,
\eea

It has been proved \cite{Vanderveld:2006rb} that the energy density is smooth only if the linear term in the above expansion is zero, i.e. $k_1=t_b^1=0$, or the Laplacian in spherical coordinates would diverge at the center.
This implies that including linear terms in the above expansions we are considering models which are not smooth at the center.
For definiteness we will present here the results of the calculations to the second order in $r$, corresponding to respectively third and fourth order for $D_L(z)$ and $mn(z)$. We will show later nevertheless that our conclusions about the existence and uniqueness of the solution are independent of the order
at which we truncate the above expansion.
 
It is important to observe that in general, up to the order at which we will consistently expand all the relevant quantities, we have eight parameters :

\be
\rho_0,\eta_0,t^b_0,t^b_1,t^b_2,k_0,k_1,k_2
\ee

or equivalently 

\be
H_0,q_0,a_0,t^b_0,t^b_1,t^b_2,k_1,k_2
\ee

where, from eq.(\ref{a0}-\ref{q0}),
 we have used the fact that $H_0,q_0,a_0$ contain the same information of $k_0,\eta_0,\rho_0$, 

There is one important question : should we regard the conformal time coordinate of the central observer $\eta_0$ as an independent parameter? Or there are some constraint to make it consistent with the age of the universe, which, if we assume the inhomogeneities to be only local, should be approximatively the same as the one estimated in $\Lambda CDM$ models?

As observed in \cite{Celerier:2009sv} since LTB solutions correspond to a pressureless spherically symmetric matter distribution, they cannot be extended to the earliest stages of cosmological history, implying that $\eta_0$ is related to the time when this kind of inhomogeneities arise, more than to the age of the Universe, and since it is re-absorbed in the definition of $H_0,q_0,a_0$ given in eq.(\ref{a0}-\ref{q0}), it is correct to consider it as an independent parameter.

As we will show later, observable do not depends on $a_0$, so it should not be considered a really independent parameter.
This is a consequence of the fact that $H_0,q_0,a_0$ are invariant under changes of parameters which preserve $\sqrt{k_0}\eta_0$.

We get six constraints from the expansion of $mn(z)$ and $D_L(z)$ respectively to fourth and third order, so in principle without imposing the smoothness conditions $t^b_1=k_1=0$ we should be able to solve the IP of locally mapping the two observable of a LTB model to any given $\Lambda CDM$ model, even taking into account the fact that, as we will show later, $mn(z)$ and $D_L(z)$ don't depend on $t^b_0$, and the solution should be unique. 

It is also clear that a simple preliminary argument based on counting the number of independent parameters implies that the IP cannot be solved if the matter distribution is smooth at the center, i.e. if $t^b_1=k_1=0$, since we would have six constraints and only four truly independent parameters, and this was shown more in detail in \cite{Romano:2009ej}.

After re-expressing the results in terms of $H_0,q_0,a_0$ and $X=\arctan{\sqrt{2 q_0-1}}$ we get:
\bea
\eta(z) &=&  \eta_0 +\eta_1 z+\eta_2 z^2 \\
r(z) &=&  r_1 z+r_2 z^2+r_3 z^3\\
\eta_1 &=& \frac{2 k_1 (q_0+1) \sqrt{2 q_0-1} X-(2 q_0-1)
   (2 a_0^3 H_0^3 (1-2 q_0)^2+2 a_0^2 H_0^3 (1-2 q_0)^2 t^b_1+3
   k_1)}{2 a_0^4 H_0^4 (2 q_0-1)^3} \nonumber\\
\eta_2 &=& \frac{1}{{4 a_0^7 H_0^7 (1-2 q_0)^4}}\bigg[-\sqrt{2 q_0-1} X (2 a_0^3 H_0^3
   k_1 (14 q_0^3-3 q_0^2-1)+\nonumber \\
   &&-4 a_0^2 H_0^2 (2 q_0-1) (2 H_0 k_1 (1-2
   q_0) q_0 t^b_1+k_2 (q_0+1))+k_1^2 (11 q_0+5))+ \nonumber\\ 
 && +(2 q_0-1) (2 a_0^6 H_0^6 (q_0+1) (2 q_0-1)^3+2 a_0^5 H_0^6 (2 q_0-1)^3 (3 q_0+1) t^b_1+ \nonumber\\
 &&+2 a_0^4 H_0^5 (2 q_0-1)^3 (H_0 (2 q_0-1) {t^b_1}^2-2t^b_2)+a_0^3 H_0^3 k_1 (26 q_0^2-11 q_0-1)+ \nonumber\\
 &&+2 a_0^2 H_0^2 (2q_0-1) (2 H_0 k_1 (2 q_0-1) t^b_1-3 k_2)+9 k_1^2)+6 k_1^2 q_0^2 
 X^2\bigg]\nonumber \\
r_1 & = & \frac{1}{a_0 H_0} \nonumber\\
r_2& = &\frac{\sqrt{2 q_0-1} \left(-a_0^3 H_0^3 \left(4 q_0^3-3 q_0+1\right)-2 a_0^2 H_0^3
   (1-2 q_0)^2 q_0 t^b_1-5 k_1 q_0+k_1\right)+6 k_1 q_0^2 X}{2 a_0^4 H_0^4 (2 q_0-1)^{5/2}} \nonumber \\
r_3& = &\frac{1}{8 a_0^7
   H_0^7 (2 q_0-1)^{9/2}}\bigg[-6 (2 q_0-1) q_0^2 X (8 a_0^2 H_0^3
   k_1 (2 q_0-1) (a_0 q_0+(2 q_0-1) t^b_1)+ \nonumber \\
 &&+  4 a_0^2 H_0^2 k_2 (1-2 q_0)+17 k_1^2)+(2 q_0-1)^{3/2} (4 a_0^6 H_0^6 (2 q_0-1)^3 (q_0^2+1)+16 a_0^5 H_0^6 q_0^2 (2 q_0-1)^3 t^b_1+ \nonumber \\
 &&+8 a_0^4 H_0^5
   q_0 (2 q_0-1)^3 (H_0 (2 q_0-1) {t^b_1}^2-t^b_2)+4 a_0^3 H_0^3
   k_1 (12 q_0^3-2 q_0^2-4 q_0+1)+\nonumber \\
 && +4 a_0^2 H_0^2 (2 q_0-1) (6 H_0
   k_1 q_0 (2 q_0-1) t^b_1-5 k_2 q_0+k_2)+k_1^2 (43 q_0-5))+\nonumber\\
  && +72k_1^2 \sqrt{2 q_0-1} q_0^3 X^2\bigg] \nonumber
\eea

We can then calculate the luminosity distance :
\bea
D_L(z)&=&(1+z)^2r(z)a(\eta(z),r(z))=D_1 z+D_2 z^2+D_3 z^3 + . .\\
D_1&=&\frac{1}{H_0}\nonumber\\
D_2&=&\frac{1}{{2 a_0^3 H_0^4 (2 q_0-1)^{5/2}}}\bigg[\sqrt{2 q_0-1} (-a_0^3 H_0^3 (1-2 q_0)^2 (q_0-1)-2 a_0^2 H_0^3 (4
   q_0^3-3 q_0+1) t^b_1-9 k_1 q_0)+\nonumber\\
   &&+6 k_1 q_0 (q_0+1) X\bigg]\nonumber\\
D_3&=&\frac{}{{4 a_0^6
   H_0^7 (2 q_0-1)^{11/2}}}\bigg[-3 (2 q_0-1) q_0 X (4 a_0^2 H_0^3
   k_1 (1-2 q_0)^2 q_0 (2 a_0 q_0+4 q_0 t^b_1+t^b_1)+\nonumber \\
   &&-4 a_0^2 H_0^2
   k_2 (4 q_0^3-3 q_0+1)+k_1^2 (50 q_0^2+7 q_0-7))+(2
   q_0-1)^{3/2} (2 a_0^6 H_0^6 (1-2 q_0)^4 (q_0-1) q_0+\nonumber\\
   &&+8 a_0^5 H_0^6 (1-2
   q_0)^4 q_0^2 t^b_1+4 a_0^3 H_0^3 k_1 (1-2 q_0)^2 q_0 (5 q_0-1)+\nonumber\\
   &&+2a_0^2 H_0^2 (1-2 q_0)^2 (H_0 k_1 (20 q_0^2+q_0-1) t^b_1-9
   k_2 q_0)+2 H_0^5 (a_0-2 a_0 q_0)^4 (H_0 q_0 (4 q_0+1)
   {(t^b_1)}^2+\nonumber\\
   &&-2 (q_0+1) t^b_2)+9 k_1^2 q_0 (11 q_0-4))+18 k_1^2 \sqrt{2
   q_0-1} (4 q_0+1) q_0^3 X^2\bigg]  \nonumber
\eea

The effects of inhomogeneities are showing already from the second order, because we are not setting the smoothness conditions  $t^b_1=k_1=0$.

From the definition of $mn(z)$ and the equation for the energy density we can write

\be
4\pi mn(z) dz = \rho d^3 V=\frac{8\pi M'}{\sqrt{1-k(r)r^2}}dr
\ee

from which by using $dr=(dr/dz)dz$ we get

\be
mn(z)=\frac{2 M'(r(z))}{\sqrt{1-k(r(z))r(z)^2}}\frac{dr(z)}{dz}=\frac{\rho_0 r(z)^2}{\sqrt{1-k(r(z))r(z)^2}}\frac{dr(z)}{dz}
\ee

where in the last equation we have used the FLRW gauge condition $M(r)=\rho_0 r^3/6$, which allows to calculate $mn(z)$ directly from $r(z)$.

We finally get:
\bea
mn(z)&=&mn_2 z^2+mn_3 z^3 +mn_4 z^4 \\
mn_2 &=&\frac{6 q_0}{H_0} \\
mn_3 &=&-\frac{2}{{a_0^3 H_0^4 (2 q_0-1)^{5/2}}}\bigg[3 q_0 (2 \sqrt{2 q_0-1} (a_0^3 H_0^3 (4 q_0^3-3 q_0+1)+2
   a_0^2 H_0^3 (1-2 q_0)^2 q_0 t^b_1+4 k_1 q_0)+\nonumber \\
   &&-12 k_1 q_0^2 X-k_1 q_0^2 \sin (4 X))\bigg]\\
mn_4 &=&-\frac{2}{32 a_0^6 H_0^7 (2 q_0-1)^5} \bigg[3 q_0 (120 q_0^2 X (\sqrt{2 q_0-1}
   (4 a_0^3 H_0^3 k_1 (1-2 q_0)^2 (3 q_0+1)+\nonumber \\
   &&-4 a_0^2 (H_0-2 H_0
   q_0)^2 (2 H_0 k_1 (1-3 q_0) t^b_1+k_2)+k_1^2 (50 q_0-17))-2 k_1^2
   q_0^2 \sin (4 X))+\nonumber \\&&
   -4 (2 q_0-1) (2
   a_0^6 H_0^6 (1-2 q_0)^4 (15 q_0^2+14 q_0+13)+40 a_0^5 H_0^6 (1-2
   q_0)^4 q_0 (3 q_0+1) t^b_1+\nonumber \\
   &&+20 a_0^3 H_0^3 k_1 (1-2 q_0)^2 (11
   q_0^2+6 q_0-2)+20 a_0^2 H_0^2 (1-2 q_0)^2 (2 H_0 k_1 q_0 (11
   q_0-4) t^b_1+\nonumber \\
   &&-5 k_2 q_0+k_2)+40 H_0^5 q_0 (a_0-2 a_0 q_0)^4
   (H_0 (3 q_0-1) {(t^b_1)}^2-t^b_2)+5 k_1^2 (136 q_0^2-73
   q_0+7))+\nonumber \\
   &&-1440 k_1^2 (3 q_0-1) q_0^3 X^2)\bigg] \nonumber
\eea


On the contrary to the smooth case, it can be seen that the effects of inhomogeneities show already from the third order, and while also the fifth order has been calculated \cite{Romano:2009qx}
, we don't report it here since we don't need it.

A final important point is that the two observables don't depend on $t^b_0$, which implies that we are left with six parameters to determine in order to solve the IP, while we will have six equations for the coefficient of the expansions.

\section{Calculating $D_L(z)$ and $mn(z)$ for $\Lambda CDM$ models.}
The metric of a $\Lambda CDM$ model is the FLRW metric, a special case of LTB solution, where :
\bea
\rho_0(r)&=&\rho_0\\
k(r)&=&0 \\
t_b(r) &=&0 \\
a(t,r)&=&a(t)
\eea

In this section we will calculate independently the expansion of the luminosity distance and the redshift spherical shell mass for the case of a flat $\Lambda CDM$.

We can also use these formulas to check the results given in the previous section, since in absence of cosmological constant and inhomogeneities they should coincide.

Here we will denote all relevant physical quantities with an upper script $\Lambda$ to avoid confusion and clearly distinguish them from the quantities  introduced previously for LTB models.

One of the Einstein equation can be expressed as:

\bea
H^{\Lambda}(z)&=&H^{\Lambda}_0\sqrt{\Omega_m{\left(\frac{a^{\Lambda}_0}{a}\right)}^3+\Omega_{\Lambda}}=H^{\Lambda}_0\sqrt{\Omega_m{(1+z)}^3+\Omega_{\Lambda}}
\eea

We can then calculate the luminosity distance using the following relation, which is only valid assuming flatness:

\bea
D^{\Lambda}_L(z)=(1+z)\int^z_0{\frac{d z'}{H^{\Lambda}(z')}}=D^{\Lambda}_1 z+D^{\Lambda}_2 z^2+D^{\Lambda}_3 z^3+ . . .
\eea

From which we can get:
\bea
D^{\Lambda}_1&=&\frac{1}{H^{\Lambda}_0}\\
D^{\Lambda}_2&=&\frac{4 \Omega_{\Lambda}+\Omega_{M}}{4 H^{\Lambda}_0}\\
D^{\Lambda}_3&=&\frac{-10 \Omega_{\Lambda} \Omega_{M}-\Omega_{M}^2}{8 H^{\Lambda}_0}
\eea

It is convenient to re-express the above coefficient in terms of the two observables $H^{\Lambda}_0,q^{\Lambda}_0$
\bea
D^{\Lambda}_1&=&\frac{1}{H^{\Lambda}_0} \\
D^{\Lambda}_2&=&\frac{1-q^{\Lambda}_0}{2 H^{\Lambda}_0} \\
D^{\Lambda}_3&=&\frac{3 (q^{\Lambda}_0)^2+q^{\Lambda}_0-2}{6 H^{\Lambda}_0}
\eea

where we have used the following relations
\bea
\Omega_L+\Omega_M&=&1 \\
\Omega_M&=&\frac{2 q^{\Lambda}_0+2}{3}
\eea

It should be underlined here that $H^{\Lambda}_0,q^{\Lambda}_0,a^{\Lambda}_0$ appearing in this formulas are not the same as the ones defined in the previous section for LTB models.

For calculating $mn(z)$ we first need $r(z)$, which can be obtained from the 
radial null geodesic equation which in this case takes the simplified form
\be
\frac{d r^{\Lambda}(z)}{d z}=\frac{1}{(1+z)\dot{a^{\Lambda}}}=\frac{1}{a^{\Lambda}_0 H^{\Lambda}}
\ee

and after integration we get

\bea
r^{\Lambda}(z)&=&\frac{1}{a^{\Lambda}_0}\int^z_0\frac{d z'}{H^{\Lambda}(z')}= \\
& &\frac{z}{a^{\Lambda}_0 H^{\Lambda}_0}-\frac{(q^{\Lambda}_0+1) z^2}{2 a^{\Lambda}_0 H^{\Lambda}_0}+\frac{\left(3 (q^{\Lambda}_0)^2+4 q^{\Lambda}_0+1\right) z^3}{6 a^{\Lambda}_0
   H^{\Lambda}_0}+O\left(z^4\right)
\eea


We can now calculate $mn^{\Lambda}(z)$
\bea
\rho_0&=&3 (a^{\Lambda}_0)^3 \Omega_M (H^{\Lambda}_0)^2 \\
mn^{\Lambda}(z)&=&mn^{\Lambda}_2 z^2+mn^{\Lambda}_3z^3+mn^{\Lambda}_4 z^4=\rho_0 r^{\Lambda}(z)^2\frac{dr^{\Lambda}(z)}{dz}=\\
&&\frac{2(q^{\Lambda}_0+1) z^2}{ H^{\Lambda}_0}-\frac{4 (q^{\Lambda}_0+1)^2 z^3}{ H^{\Lambda}_0}+\frac{5 (q^{\Lambda}_0+1)^2 (9 q^{\Lambda}_0+5) z^4}{6
   H^{\Lambda}_0}+O\left(z^5\right)
\eea

We can check the consistency between these formulae and the one derived in the case of LTB without cosmological constant by setting:
\bea
k_1=k_2=t_1^b=t_2^b=0\\
q_0=1/2
\eea

which corresponds to the case in which $\Omega_M=1$.

It is important to mention again that $a^{\Lambda}_0,q^{\Lambda}_0,H^{\Lambda}_0$ defined in this section are in general different from the ones defined in the previous section for LTB models.

\section{Local solution of the inversion problem}
In this section we will denote with an upper script $\Lambda$ all the relevant quantities referred to a $\Lambda CDM$ model, including the coefficients of the redshift expansion for $D_L(z)$ and $mn(z)$.
 
In order to solve he inversion problem to the second order in $r$ we need to solve the following system of six equations equations:
\bea
D_i^{\Lambda}=D_i& 1\leq i\leq 3\\
mn_j^{\Lambda}=mn_j& 2\leq j\leq 4 
\eea

Including higher orders in the expansion would not change our conclusions since if we include a new order for $k_n,t^b_n$ we would than have two new parameters to solve for and two new conditions coming from the matching of the coefficient $\{D_{n+1},mn_{n+2}\}$ in the redshift expansion.

We can first solve the system of two equations
\bea
D_1^{\Lambda}=D_1\\
mn_2^{\Lambda}=mn_2 
\eea

which gives the simple solution:

\bea
H_0=H^{\Lambda}_0 \\
q_0=\frac{1+q_0^{\Lambda}}{3} \label{q0NSsol}
\eea

Since $q_0^{\Lambda}\geq -1$, the above equation implies $q_0>0$, so that the positivity condition implied by eq.(\ref{q0}) is respected, and the inversion problem is possible.
We are left with four parameters $\{k_1,k_2,t^b_1,t^b_2\}$ to determine, and since we have only four more independent equations, the solution should be unique, but we will report it in detail in a separate work, while here we are interested in proving the existence of such a solution.

Another important consequence of our calculation is that it is impossible to solve the inversion problem for both $mn(z)$ and $D_L(z)$ setting one the two functions $k(r)$ or $t^b(r)$ to zero, even allowing not smooth matter profiles, since we would then have less parameters to solve for than equations.

On the other side as long as we are trying to solve the inversion problem for a single observable, we would always have more free parameters than equations, even imposing the smoothness conditions, meaning that in this kind of IP there is not a unique local solution.
We could use this freedom for example to smooth the density profile. 
We will further investigate this a in a future work, where we will study the local solutions to the IP for $D_L(z)$ and $mn(z)$ separately and for both the observables at the same time. 
In the smooth case \cite{Romano:2009ej} there is no solution of the IP for both $D_(z)$ and $mn(z)$, since there are not enough free parameters to solve the IP equations.
In the smooth single observable IP for $D_L(z)$ we obtain instead $q_0=q_0^{\Lambda}$, which implies the well known fact that negative $q_0^{\Lambda}$ cannot be reproduced by LTB models with smooth matter distribution \cite{Vanderveld:2006rb}, since $q_0$ is always positive.
This single observable smooth case IP solution is completely different from the not smooth multiple observable case in eq.(\ref{q0NSsol}) since they correspond to different types of IP, i.e. they satisfy different set of equations. For this reason the single observable smooth solution cannot be obtained from the multiple observable solutions in the $k_1=t^b_1=0$ limit because they are different sets of solutions, and in fact the IP has no solution in the multiple observable smooth case.

It should be also mentioned that since according to our analysis we are left with no free parameter after solving the IP for both $mn(z)$ and $D_L(z)$, it is highly probable that additional independent observables should make possible for a central observer to distinguish without any ambiguity between matter dominated LTB and $\Lambda CDM$ models.

\section{Do we need a void to explain luminosity distance observations? }
So far we have focused on the possibility to solve the IP for both $mn(z)$ and $D_L(z)$ in presence of a centrally not smooth spherically symmetric matter distribution, showing how the on the contrary to the smooth case a solution should exist, as far as $H_0$ and $q_0$ are concerned, giving they explicit relation with $q_0^{\Lambda}$ and $H_0^{\Lambda}$:

\be
H_0^{LTB}=H^{\Lambda}_0 \quad,\quad q_0^{LTB}=\frac{1+q_0^{\Lambda}}{3}.
\ee

It actually turns out that the solution is unique, since after introducing the two following dimensionless parameters :
\bea
K_1=(a_0 H_0)^{-3}k_1 \quad,\quad K_2= (a_0 H_0)^{-4} k_2\\
T_b^1=a_0^{-1} t_b^1 \quad,\quad T_b^2=a_0^{-1}(a_0 H_0)^{-1} t_b^2 
\eea
we can eliminate $a_0$ in the formulae for the observables, which means that, as expected, they do not depend on $a_0$.
For example we can get :
\bea
D_2&=&\frac{6 K_1 q_0 (q_0+1) \sqrt{2 q_0-1} X-(2 q_0-1) \left(q_0 (9 K_1-6
   T_b^1+5)+q_0^3 (8 T_b^1+4)-8 q_0^2+2 T_b^1-1\right)}{2 H_0 (2 q_0-1)^3} \nonumber \\ 
   &&
\eea 

This can be seen as a consequence of the fact that $\eta_0$ and $k_0$ always appear in the form $\sqrt{k_0}\eta_0$  in the definition of $a_0,q_0,H_0$, implying that they do not really correspond to two independent degrees of freedom of the problem, since physical quantities are invariant under changes of parameters which preserve $\sqrt{k_0}\eta_0$.
Since $t_b^0$ never appears in the expansion of the observables, we are left with only six independent parameters and six independent constraints, implying that there is at most only on independent solution to the IP

A detailed analysis of the solution is left to a future work, while here we will consider the consequences of our results on fitting the luminosity distance data.
Let us start from considering the constraints coming from the first two terms:
\be
D_0=D_0^{\Lambda} \quad,\quad D_1=D_1^{\Lambda} \label{eqDD}
\ee
which imply that out of the four independent parameters $H_0,q_0,K_1,T_1,$ two of are left free, so the IP is degenerate, even considering a homogeneous bang function, i.e. setting $T_1$. This is confirmed by the fact that several different solutions have been prosed in the literature with $t_b(r)=0$. 
Going to a higher order would increase the degeneracy of the solution, since we would then only have one extra constraint coming from $D_n$ and two additional parameters to determine, $K_n,T_n$.

To give a more definite answer to the question posed in the title of this section, we should actually look for the best fit parameters and then calculate the corresponding density profile, but this would go beyond the scope of the present paper which is to prove that a centrally not smooth matter distribution can mimick the cosmological constant for both $mn(z)$ and $D_(z)$.

We can nevertheless observe, based on the recent numerical work by \cite{Celerier:2009sv} and our analysis performed so far, that in presence of a not homogeneous bang function there exist solutions of the IP for both $mn(z)$ and $D_L(z)$ which correspond to overdense regions.
This solutions are obviously valid even in the case in which we consider only $D_L(z)$, implying that as far as $D_L(z)$ is considered, both overdense or underdense regions can fit the data if we include inhomogeneous bang functions.

In the homogeneous bang case $t_b(r)=0$, only solutions corresponding to a central void are known, and we leave to a future work a detailed study of this case based on the low redshift expansion we have derived, in order to determine if this is the only type of possible models, but based on physical arguments about the dynamics of void expansion we expect that the dimming of supernovae light can only be explained by an underdense region for the homogeneous bang case.

One last important comment is that so far we have focused on reproducing the coefficient of the redshift expansion of the observables corresponding to the best fit flat $\Lambda CDM$ models, but this kind of approach is not completely rigorous since these models depend only on the two independent parameters $H_0,\Omega_M$, while LTB models have a higher number of parameters, implying that they could actually provide an even better fit of experimental data.

\section{Conclusion}

We have calculated for a central observer in a LTB space without cosmological constant the low redshift expansion for  the luminosity distance and the redshift spherical shell mass respectively to third and fourth order. We have then derived the same observables for $\Lambda CDM$ models and shown how it is in principle possible to locally mimick a $\Lambda CDM$ model with a LTB matter dominated universe if the conditions for a centrally smooth matter distribution are not imposed, i.e. if we don't set $k_1=t^b_1=0$, and the solution is  unique.

Another important consequence of our calculation is that it is impossible to solve the IP for both $mn(z)$ and $D_L(z)$ setting to zero one the two functions $k(r)$ or $t^b(r)$, even allowing non smooth matter density profiles, since we would then have more independent equations to solve than parameters to determine.

For the case of the inversion problem for a single observable we have an even wider space of solutions and we could try to use this freedom to impose the conditions for a smooth matter profile. It is in fact possible to find a smooth matter distribution able to reproduce the best fit $\Lambda CDM$ redshift expansion for the luminosity distance if the bang function is not homogeneous, and we will investigate this in detail in a separate work.

It should be also mentioned that since according to our analysis we are left with no more free parameters after solving the IP for $mn(z)$ and $D_L(z)$, it is highly probable that additional independent observables would make possible for a central observer to distinguish without any ambiguity between matter dominated LTB and $\Lambda CDM$ models.

We will further investigate this in a future work, where we will study in detail the local solutions to the IP for $D_L(z)$ and $mn(z)$ separately, for both the observables at the same time, and for other relevant observables, determining when overdense or underdense regions are possible solutions.

One last important comment is that we have focused on reproducing the coefficient of the redshift expansion of the observables corresponding to the best fit flat $\Lambda CDM$ models, but this kind of approach is not completely rigorous since these models depend only on the two independent parameters $H_0,\Omega_M$, while LTB models have a higher number of parameters, implying that they could actually provide an even better fit of experimental data. 
From an observational point of view it would in fact be more important to confront directly with actual experimental data rather than reproduce the best fit theoretical $\Lambda CDM$ model, so our conclusions should be considered keeping this limitation in mind, and we leave to a future work such a direct analysis of experimental data.

\begin{acknowledgments}
I thank A. Starobinsky, M. Sasaki, M.~N.~Celerier e and A. Notari for useful comments and discussions, and J. Yokoyama for the 
for the hospitality at RESCUE. This work is supported by the Grant-in-Aid for the Global COE
Program "The Next Generation of Physics, Spun from Universality and Emergence" from MEXT.

\end{acknowledgments}

\end{document}